\documentclass{iopart}

\usepackage{cite}
\expandafter\let\csname equation*\endcsname\relax
\expandafter\let\csname endequation*\endcsname\relax
\usepackage{amsmath}
\usepackage{amssymb}
\usepackage{amsthm} 
\usepackage{amsfonts}
\usepackage{color}
\usepackage{graphicx}

\begin{document}

\title{Spectral singularity in composite systems and simulation of a
resonant lasing cavity}
\author{X. Z. Zhang$^1$, G. R. Li$^2$, and Z. Song$^2$\footnote[1]{\textsf{songtc@nankai.edu.cn}}}

\address{$^1$College of Physics and Materials Science, Tianjin Normal University, Tianjin
300387, China}
\address{$^2$School of Physics, Nankai University, Tianjin
300071, China} 

\begin{abstract}
We investigate herein the existence of spectral singularities (SSs) in composite
systems that consist of two separate scattering centers A and B
embedded in one-dimensional free space, with at least one scattering center being non-Hermitian. We show that such composite systems have an SS at $k_{c}$ if the
reflection amplitudes $r^{A}\left( k_{c}\right) $ and $r^{B}\left(
k_{c}\right) $ of the two scattering centers satisfy the condition $r_{\mathrm{R}%
}^{A}\left( k_{c}\right) r_{\mathrm{L}}^{B}\left( k_{c}\right)
e^{i2k_{c}\left( x_{B}-x_{A}\right) }=1$. We also extend the condition to the system with multi-scattering centers. As an application, we
construct a simple system to simulate a resonant lasing cavity.
\end{abstract}

\maketitle

\pagenumbering{arabic}

\section{Introduction}

Non-Hermitian Hamiltonians have long been used to describe open quantum systems in fields
ranging from condensed-matter physics to nuclear physics and in particular, quantum optics \cite%
{Moiseyev,Muga,Rotter}. Since the pioneering work of Bender and Boettcher \cite%
{Bender1998}, which opened a new branch of quantum mechanics, a series of non-Hermitian Hamiltonians has been proposed combining parity ($%
\mathcal{P}$) and time-reversal symmetry ($\mathcal{T}$) and real spectra. This discovery opened a new branch of quantum mechanics. Among the many interesting peculiarities of the non-Hermitian Hamiltonians, that have received significant attention is the presence of exceptional points and spectral singularities
(SSs) in their spectra, that prevent the system from being diagonalised \cite%
{MostafazadehPRL2009,Andrianov,MostafazadehJPA,MostafazadehPRA83,MostafazadehPRA2012}.
Exceptional points refer to the coalescence of two or more eigenstates
together with their eigenvalues when the spectrum is discrete. Their
physical relevance has been investigated in several contexts \cite%
{Kato,Berry,Kemp,Naimark,Schwartz,Ljance,Samsonov,Dehnavi,Heiss,LonghiPRB},
and the corresponding physical realizations have been experimentally
demonstrated. SSs have a nature similar to that of exceptional points, but within the continuous spectrum of a non-Hermitian operator. Furthermore, SSs appear as zeros of the Wronskian of Jost solutions and as divergences of the resolvant operator. In 2009, Mostafazadeh showed that SSs in non-periodic complex
potentials correspond to resonances with vanishing spectral width \cite%
{MostafazadehPRL2009}. This stimulated numerous investigations into the physical meaning and relevance of SSs in the framework of wave scattering, in which SSs exhibit fascinating phenomena such as lasing or antilasing in optics \cite%
{MostafazadehPRA83,Longhiphysics,LonghiPRA1,LonghiPRA2,LonghiPRL,Chong,Wan,Ge,Stone,Sarisaman,Zhang}, particle-pair annihilation in many body systems \cite{Li}, and flux-induced asymmetric transmission and transmission phase lapse \cite%
{ZhangG,LiXQ}. Therefore, a theoretical search for systems with SSs is imperative.

With this motivation, we investigate herein the existence of an SS in a composite system consisting of two separate scattering centers. The existence of an
SS requires the reflection amplitudes $r^{A}\left( k_{c}\right) $ and $%
r^{B}\left( k_{c}\right) $ of the two scattering centers to satisfy the
condition $r_{\mathrm{R}}^{A}\left( k_{c}\right) r_{\mathrm{L}}^{B}\left(
k_{c}\right) e^{i2k_{c}\left( x_{B}-x_{A}\right) }=1$, which provides a
simple way of constructing such the systems with SS at any energy in demand. These systems can also be used to simulate a resonant lasing cavity.
On the other hand, based on the theorem proposed by Mostafazadeh \cite%
{MostafazadehPRL2009}, we can also extend the results to the systems with
multi-scattering centers, which can pave the way to the experimental demonstration of diverse systems with SSs.

This paper is organized as follow. In Sec. \ref{sec_gain}, we present a
condition about the existence of the SSs in composite systems, and then
apply it to the continuous and discrete systems. Sec. \ref{sec_simulation}
is devoted to simulating the lasing process through the dynamics of a
half-infinite non-Hermitian continuous system at an SS.
Finally, we give a summary in Sec. \ref{sec_summary}.

\section{Gain scattering center and SS}

\label{sec_gain}

\subsection{the condition for the composite non-Hermitian system with SS}

\label{sec_secle theorem}

\begin{figure}[tbp]
\centering
\includegraphics[ bb=44 145 546 750, width=0.6\textwidth, clip]{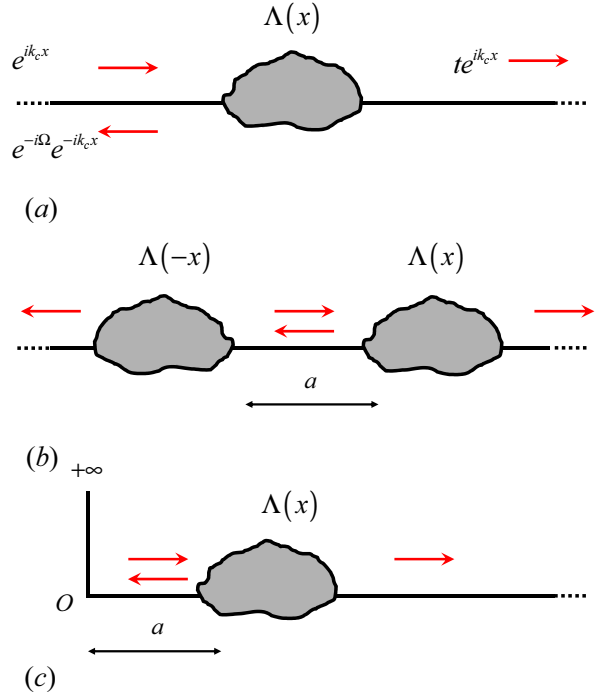}
\caption{(Color online) Schematic illustration of the condition proposed in
the Sec. \protect\ref{sec_secle theorem}. (a) Left-incident solution for
a scattering center $\Lambda \left( x\right) $. (b) Composite scattering
system consisting of scattering center $\Lambda \left( x\right) $ and its
complementary scattering center $\Lambda \left( -x\right) $ seperated by distance $a$. The left incident and reflection plane waves of $\Lambda \left( x\right) $
are the right reflection and incident plane waves of $\Lambda \left(
-x\right) $, respectively, which satisfy the Eq. (\protect\ref{MC}). The
corresponding SS of the system can be identified by the form of the solution
far from the scattering center, which represents either a reflectionless
absorption (antilasing) or a self-sustaining emission (lasing). Note that, for any given left-incident solution of a non-Hermitian scattering center $\Lambda \left( x\right) $, one can choose $\Lambda \left( -x\right) $
to satisfy the Eq. (\protect\ref{MC}) to realize a composite system with an SS.
(c) Illustrative example composed of an infinite ending potential and a
scattering center $\Lambda \left( x\right) $. The scattering center $\Lambda
\left( x\right) $ can be deemed as a complementary scattering center of the
infinite ending potential with $r_{\mathrm{R}}^{A}\left( k_{c}\right)
e^{i2k_{c}a}=r_{\mathrm{L}}^{B}\left( k_{c}\right) =-1$.}
\label{fig1}
\end{figure}

We consider a composite scattering center (CSC), which consists of two
separate scattering centers $A$ and $B$, located at $x_{a}$ and $x_{b}$,
respectively. A CSC can be characterized by a transfer
matrix $M\left( k\right) $, which can be obtained from the product of the
transfer matrices $M_{A}\left( k\right) $\ and $M_{B}\left( k\right) $\ of
the individual scattering centers $A$ and $B$. For a single center $A$ or $B$%
, located at the origin, the corresponding Jost solutions defined
in terms of their asymptotic behavior are%
\begin{equation}
\psi _{\mathrm{R}}^{\rho }\left( x\right) =\left\{
\begin{array}{cc}
e^{-ikx}+r_{\mathrm{R}}^{\rho }\left( k\right) e^{ikx}, & \left(
x\rightarrow \infty \right) \\
t_{\mathrm{R}}^{\rho }\left( k\right) e^{-ikx}, & \left( x\rightarrow
-\infty \right)%
\end{array}%
\right. ,
\end{equation}%
and%
\begin{equation}
\psi _{\mathrm{L}}^{\rho }\left( x\right) =\left\{
\begin{array}{cc}
e^{ikx}+r_{\mathrm{L}}^{\rho }\left( k\right) e^{-ikx}, & \left(
x\rightarrow -\infty \right) \\
t_{\mathrm{L}}^{\rho }\left( k\right) e^{ikx}, & \left( x\rightarrow \infty
\right)%
\end{array}%
\right. ,
\end{equation}%
where $r_{\mathrm{\sigma }}^{\rho }\left( k\right) $, $t_{\mathrm{\sigma }%
}^{\rho }\left( k\right) $ $\left( \rho =A,B\text{ and }\mathrm{\sigma =L,R}%
\right) $ are the reflection and transmission amplitudes. This results in%
\begin{equation}
M_{\rho }\left( k\right) =\left(
\begin{array}{cc}
t_{\mathrm{L}}^{\rho }-r_{\mathrm{R}}^{\rho }r_{\mathrm{L}}^{\rho }/t_{%
\mathrm{R}}^{\rho } & e^{-2ikx_{\rho }}r_{\mathrm{R}}^{\rho }/t_{\mathrm{R}%
}^{\rho } \\
-e^{2ikx_{\rho }}r_{\mathrm{L}}^{\rho }/t_{\mathrm{R}}^{\rho } & 1/t_{%
\mathrm{R}}^{\rho }%
\end{array}%
\right) ,
\end{equation}%
which relates to the initial state and the final state of a physical system
undergoing a scattering process, and%
\begin{equation}
M\left( k\right) =M_{B}\left( k\right) M_{A}\left( k\right) .
\end{equation}%
A quantity of interest are the entries of the above matrix, especially in the
form of $M_{22}\left( k\right)$, which can be expressed as
\begin{equation}
M_{22}\left( k\right) =\left[ 1-e^{i2k\left( x_{B}-x_{A}\right) }r_{\mathrm{R%
}}^{A}r_{\mathrm{L}}^{B}\right] /\left( t_{\mathrm{R}}^{A}t_{\mathrm{R}%
}^{B}\right) .
\end{equation}%
According to the theorems proposed in Ref. \cite{MostafazadehPRL2009}, the zeros
of $M_{22}\left( k\right) $ at $k_{c}$, i.e,%
\begin{equation}
r_{\mathrm{R}}^{A}\left( k_{c}\right) r_{\mathrm{L}}^{B}\left( k_{c}\right)
e^{i2k_{c}\left( x_{B}-x_{A}\right) }=1,  \label{MC}
\end{equation}%
admits an SS for the CSC. The essential condition is $\left\vert r_{\mathrm{R}%
}^{A}\left( k_{c}\right) r_{\mathrm{L}}^{B}\left( k_{c}\right) \right\vert
=1 $\ because it is always possible to adjust $x_{B}-x_{A}$ to satisfy the original CSC in Eq. (\ref{MC}). Furthermore, the condition of $\left\vert r_{\mathrm{R}%
}^{A}\left( k_{c}\right) \right\vert \left\vert r_{\mathrm{L}}^{B}\left(
k_{c}\right) \right\vert =1$ indicates that either scattering center $A$ or $%
B$ is non-Hermitian because $\left\vert r_{\mathrm{R}%
}^{A}\left( k_{c}\right) \right\vert \geqslant 1$ or $\left\vert r_{\mathrm{L%
}}^{B}\left( k_{c}\right) \right\vert \geqslant 1$ with $\left\vert t_{%
\mathrm{\sigma }}^{\rho }\left( k\right) \right\vert \neq 0$. These
results imply two significant points: (i) For given left- or
right-going scattering solutions of the gain system, one can relate it to a
composite system with an SS via the matching condition (\ref{MC}). (ii) One
can construct a system with SSs at any desired energy by modulating the
parameters of the scattering centers involved. For the sake of clarity, we
illustrate this condition in Fig. \ref{fig1} through two simple composite
systems. Note that the system shown in Fig. \ref{fig1}(c), which consists of
an infinite ending potential and scattering center $\Lambda\left( x\right) $%
, will be employed to simulate the lasing process in the Sec. \ref%
{sec_simulation}.

\subsection{illustrative examples}

\label{sec_secle} In this subsection, we apply the condition to
continuous and discrete systems to construct a composite non-Hermitian
system with an SS.

\subsubsection{SS of multi delta potentials}

\label{sec_secle 1}Consider a non-Hermitian continuous system with
double imaginary delta potentials. The Hamiltonian of which has the form%
\begin{equation}
H^{[2]}=-\frac{\partial ^{2}}{2\partial x^{2}}+i\sum_{l=1}^{2}V_{l}\delta
\left( x-x_{l}\right) .
\end{equation}%
If the delta potential is at the origin, then the reflection amplitudes
are
\begin{equation}
r_{\mathrm{R}}^{l}\left( k\right) =r_{\mathrm{L}}^{l}\left( k\right)
=V_{l}/\left( k-V_{l}\right) ,
\end{equation}%
which can be shown in the Ref. \cite{JPAAli}. Based on the matching
condition (\ref{MC}), we find that this composite system has an SS at $k_{c}$%
\ if%
\begin{equation}
k_{c}=V_{1}+V_{2}\text{, }k_{c}\left( x_{2}-x_{1}\right) =m\pi ,
\end{equation}%
where $m=0,1,2,...$. Meanwhile, the two corresponding degenerate solutions
coalesce and can be expressed as%
\begin{eqnarray}
\psi ^{1}\left( k_{c},x\right) &=&\psi ^{2}\left( k_{c},x\right) \\
&=&\left\{
\begin{array}{c}
e^{-i\left( V_{1}+V_{2}\right) x},\text{ }\left( x<x_{1}\right) \\
\cos \left( V_{1}+V_{2}\right) x+i\frac{V_{1}-V_{2}}{V_{1}+V_{2}} \\
\times \sin \left( V_{1}+V_{2}\right) x,\text{ }\left( x_{1}\leqslant
x<x_{2}\right) \\
e^{i\left( V_{1}+V_{2}\right) x},\text{ }\left( x\geqslant x_{2}\right)%
\end{array}%
\right. .  \notag
\end{eqnarray}%
%
%
%
%
%
%
%

\begin{figure}[tbp]
\centering
\includegraphics[ bb=57 69 558 777, width=0.6\textwidth, clip]{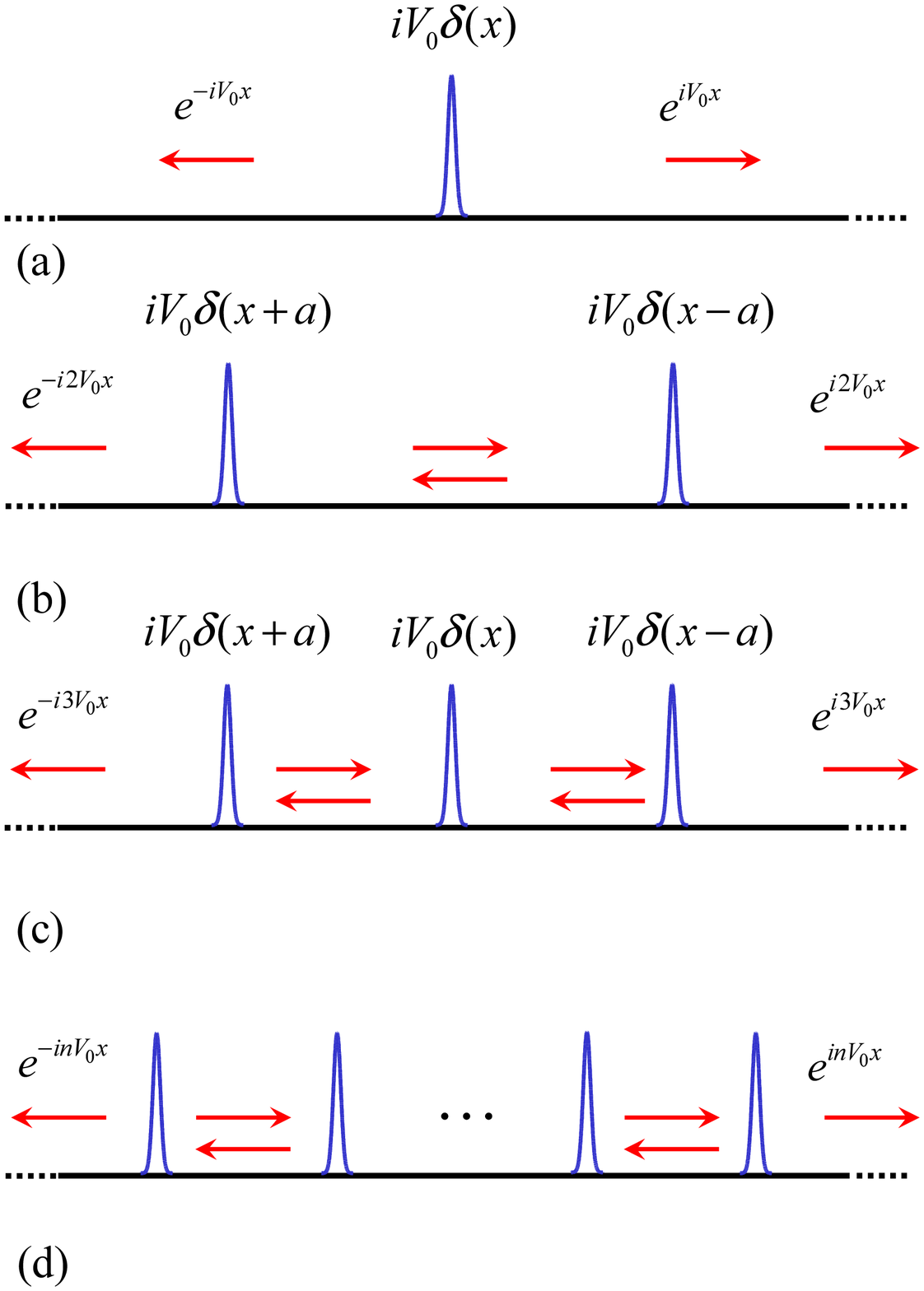}
\caption{(Color online) Schematic of the solution at an SS with Hamiltonian (%
\protect\ref{H_n}) for the case of (a) $n=1$, (b) $n=2$, (c) $n=3$, (d) $%
n=\infty $, respectively.}
\label{fig2}
\end{figure}

%
%
%
%
%
%
We note that for $m=0$, the result goes back to the single delta-potential
case. On the other hand, we can construct a set of eigenfunctions of $\left(
H^{[2]}\right) ^{\dagger }$ in terms of their asymptotic behaviour. They are
given by%
\begin{eqnarray}
\overline{\psi }^{1}\left( k_{c},x\right) &=&\overline{\psi }^{2}\left(
k_{c},x\right) \\
&=&\left\{
\begin{array}{c}
e^{i\left( V_{1}+V_{2}\right) x},\text{ }\left( x<x_{1}\right) \\
\cos \left( V_{1}+V_{2}\right) x-i\frac{V_{1}-V_{2}}{V_{1}+V_{2}} \\
\times \sin \left( V_{1}+V_{2}\right) x,\text{ }\left( x_{1}\leqslant
x<x_{2}\right) \\
e^{-i\left( V_{1}+V_{2}\right) x},\text{ }\left( x\geqslant x_{2}\right)%
\end{array}%
\right. .  \notag
\end{eqnarray}%
This implies that the existence of an SS prevents the
diagonalization of $H^{[2]}$. It is worthy pointing that the contribution of the SS to the resolution of identity
operator depends on the class of functions employed for physical states and that there is no obstruction to completeness originating
from a spectral singularity\cite{Andrianov}. Furthermore, we can also extend our analysis to the $n$-delta potential case%
\begin{equation}
H^{[n]}=-\frac{\partial ^{2}}{2\partial x^{2}}+i\sum_{l=1}^{n}V_{l}\delta
\left( x-x_{l}\right) ,  \label{H_n}
\end{equation}
and the corresponding transfer matrix of the imaginary delta potential $%
iV_{l}\delta\left( x-x_{l}\right)$ can be written as%
\begin{equation}
M_{l}\left( k\right) =\left(
\begin{array}{cc}
1+\frac{V_{l}}{k} & e^{-2ikx_{l}}\frac{V_{l}}{k} \\
-e^{2ikx_{l}}\frac{V_{l}}{k} & 1-\frac{V_{l}}{k}%
\end{array}%
\right) .
\end{equation}%
On the other hand, the multipotentials can be considered as a CSC with the transfer matrix
\begin{equation}
M\left( k\right) =\prod\limits_{l=1}^{n}M_{n+1-l}\left( k\right) .
\end{equation}%
Taking $k_{c}\left( x_{l+1}-x_{l}\right) =m\pi $ ($m$ is an integer), $%
M\left( k_{c}\right) $ reduces to
\begin{equation}
M\left( k_{c}\right) =1+\frac{\left( \sigma _{z}+i\sigma _{y}\right) }{k_{c}}%
\sum_{j=1}^{n}V_{j},
\end{equation}%
where $\sigma _{z}$ and $\sigma _{y}$\ are Pauli matrices. Obviously, the
condition of $k_{c}=\sum_{j=1}^{n}V_{j}$ yields
\begin{equation}
M_{22}\left( k_{c}\right) =0,
\end{equation}%
which corresponds to an SS of the system at $k_{c}$. We schematically
illustrate the solution at the SS for the Hamiltonian (\ref{H_n}) in Fig. \ref%
{fig2}, in which the delta potentials have equal strength. Note that the solution between two adjacent delta potentials satisfies condition Eq. (\ref{MC}), which admits an SS in the composite system. And
the corresponding $k_{c}$ is proportional to the sum of the strengths $V_{0}$ of each delta potential (i.e., $nV_{0}$).

\subsubsection{SS of a non-Hermitian discrete system possessing a
non-Hermitian and a Hermitian scattering centers}

\begin{figure}[tbp]
\begin{center}
\includegraphics[ bb=40 132 524 722, width=0.6\textwidth, clip]{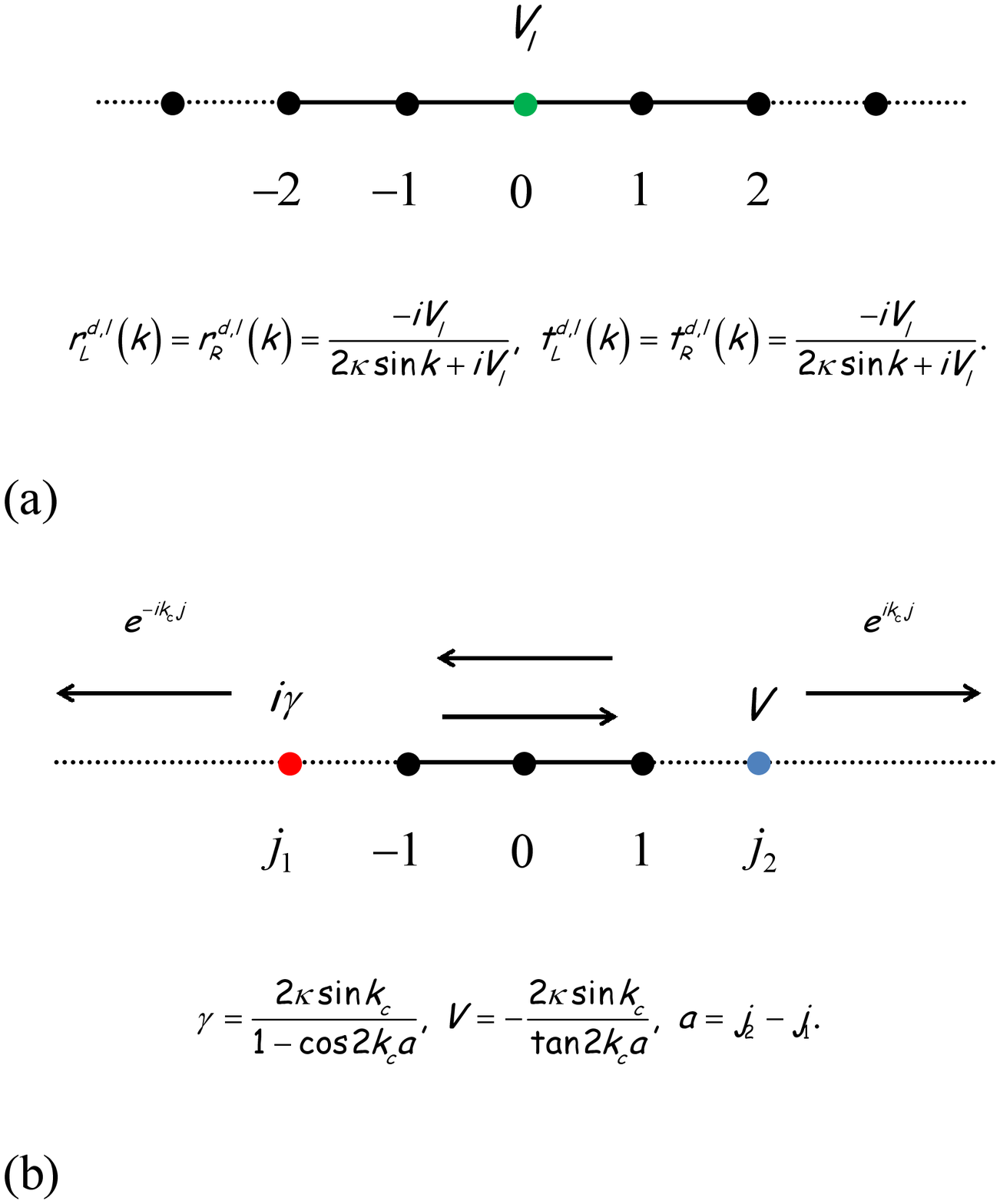}
\caption{(Color online) Schematic depiction of the considered discrete composite non-Hermitian system. (a) Scattering system with a single potential at the origin. The scattering solution for this can be used to construct
a composite non-Hermitian scattering system with an SS. (b)
Composite system at the SS. Note that, according to Eq. (\ref{potentials}),
the strength of the potentials can be adjusted to prepare a composite system with
an SS at any $k_{c}\in \left( 0,\pi \right) $ in demand.}
\label{fig3}
\end{center}
\end{figure}

\label{sec_secle 2}Our findings are not restricted to continuous systems
but would hold for the discrete systems, in which one can construct a
composite system with an SS via the method outlined in Sec. \ref{sec_secle
1}. Because the components of the CSC can be Hermitian or
non-Hermitian, we can construct such a discrete composite system by matching
the gain to the Hermitian scattering center. Towards this end, we start by considering Hamiltonian of a scattering tight-binding network with a CSC%
\begin{equation}
H_{d}=H_{0}+\sum_{l=1}^{2}H^{l}\left( j_{l}\right)
\end{equation}%
where%
\begin{eqnarray}
H_{0} &=&-\kappa \sum_{j=-\infty }^{\infty }\left( \left\vert j\right\rangle
\left\langle j+1\right\vert +\text{\textrm{H.c.}}\right) , \\
H^{l}\left( j_{l}\right)  &=&V_{l}\left\vert j_{l}\right\rangle \left\langle
j_{l}\right\vert ,
\end{eqnarray}%
represent a free tight-binding chain and the on-site potential in the
single-particle invariant space, respectively. We sketch this
Hamiltonian in Fig. \ref{fig3}. Experimentally, this Hamiltonian can be realized in
coupled optical waveguides through guiding and a inclusion of balanced gain
and loss region \cite{MusslimaniOL,Makris08,AGuo,CERuter}. To obtain the
SS of the whole system, we first investigate the solution
of the system consisting of $H_{0}$ and $H^{l}\left( 0\right) $, i.e.,%
\begin{equation}
H_{d}^{l}=H_{0}+H^{l}\left( 0\right) ,
\end{equation}%
the right- and left-going solution of the $H_{d}^{l}$ can be obtained by using the
Bethe ansatz method, which is shown in Fig. \ref{fig3}(b). The corresponding
wave functions have the form%
\begin{equation}
\psi _{\mathrm{R}}^{d,l}\left( j\right) =\left\{
\begin{array}{cc}
e^{-ikj}+r_{\mathrm{R}}^{d,l}\left( k\right) e^{ikj}, & \left( j\geqslant
0\right)  \\
t_{\mathrm{R}}^{d,l}\left( k\right) e^{-ikj}, & \left( j<0\right)
\end{array}%
\right. ,
\end{equation}%
\begin{equation}
\psi _{\mathrm{L}}^{d,l}\left( j\right) =\left\{
\begin{array}{cc}
e^{ikj}+r_{\mathrm{L}}^{d,l}\left( k\right) e^{-ikj}, & \left( j<0\right)
\\
t_{\mathrm{L}}^{d,l}\left( k\right) e^{-ikj}, & \left( j\geqslant 0\right)
\end{array}%
\right. .
\end{equation}%
Here $r_{\mathrm{\sigma }}^{d,l}\left( k\right) $, $t_{\mathrm{\sigma }%
}^{d,l}\left( k\right) $ $\left( l=1,2\text{ and }\mathrm{\sigma =L,R}%
\right) $ are the reflection and transmission amplitudes of the incident
wave with energy $E=-2\kappa \cos k$, $k\in \left[ 0,\pi \right] $. By
substituting the wave function $\psi _{\mathrm{\sigma }}^{d,l}\left(
j\right) $ into the Schr\"{o}dinger equation,%
\begin{eqnarray}
-\kappa \psi _{\mathrm{\sigma }}^{d,l}\left( j-1\right) -\kappa \psi _{%
\mathrm{\sigma }}^{d,l}\left( j+1\right) =E\psi _{\mathrm{\sigma }%
}^{d,l}\left( j+1\right) , &j\notin \left( 0\right) \text{,}& \\
-\kappa \psi _{\mathrm{\sigma }}^{d,l}\left( -1\right) -\kappa \psi _{%
\mathrm{\sigma }}^{d,l}\left( 1\right) =\left( E-V_{\rho }\right) \psi _{%
\mathrm{\sigma }}^{d,l}\left( 0\right) , &&
\end{eqnarray}%
we obtain%
\begin{eqnarray}
r_{\mathrm{L}}^{d,l}\left( k\right)  &=&r_{\mathrm{R}}^{d,l}\left( k\right) =%
\frac{-iV_{l}}{2\kappa \sin k+iV_{l}}, \\
t_{\mathrm{L}}^{d,l}\left( k\right)  &=&t_{\mathrm{R}}^{d,l}\left( k\right) =%
\frac{2\kappa \sin k}{2\kappa \sin k+iV_{l}}.
\end{eqnarray}%
For the sake of simplicity, we take $V_{1}=i\gamma $, $V_{2}=V$ to
characterize the non-Hermitian and Hermitian scattering centers of the CSC,
respectively. The corresponding eigen wave functions reduce to
\begin{equation}
\psi _{\mathrm{R}}^{d,1}\left( j\right) =\left\{
\begin{array}{cc}
e^{-ikj}+\frac{\gamma }{2\kappa \sin k-\gamma }e^{ikj}, & \left( j\geqslant
0\right)  \\
\frac{2\kappa \sin k}{2\kappa \sin k-\gamma }e^{-ikj}, & \left( j<0\right)
\end{array}%
\right. ,
\end{equation}%
\begin{equation}
\psi _{\mathrm{L}}^{d,2}\left( j\right) =\left\{
\begin{array}{cc}
e^{ikj}+\frac{-iV}{2\kappa \sin k+iV}e^{-ikj}, & \left( j<0\right)  \\
\frac{2\kappa \sin k}{2\kappa \sin k+iV}e^{-ikj}, & \left( j\geqslant
0\right)
\end{array}%
\right. .
\end{equation}%
When $\sin k<\gamma /\kappa $ $\left( \left\vert \gamma /\kappa \right\vert
\leqslant 1\right) $, the scattering center $H^{1}\left( 0\right) $ can be
deemed as a gain scattering center with $\left\vert r_{\mathrm{\sigma }%
}^{d,1}\left( k\right) \right\vert >1$ and $\left\vert t_{\mathrm{\sigma }%
}^{d,1}\left( k\right) \right\vert \neq 0$. Based on the matching condition (%
\ref{MC}), we find that this composite system has an SS at $k_{c}$, which is
determined by%
\begin{equation}
\gamma =\frac{2\kappa \sin k_{c}}{1-\cos 2k_{c}a},V=-\frac{2\kappa \sin k_{c}%
}{\tan 2k_{c}a},  \label{potentials}
\end{equation}%
where $a=j_{2}-j_{1}$ is an integer and denotes the distance between the on
site potentials $H^{1}\left( j_{1}\right) $ and $H^{2}\left( j_{2}\right) $.
This indicates that for a given scattering solution $\psi _{\mathrm{R}%
}^{d,1}\left( j\right) $ of $H_{d}^{1}$, there can exist a suitable
Hermitian potential $V$ to form a composite non-Hermitian system at the SS,
which shares the same solution in the region of $j\geqslant j_{2}$. This result
also shows, in the context of lattice model, the SS can be prepared at energy in demand.

\section{Simulation of a resonant cavity}

\label{sec_simulation}

\begin{figure}[tbp]
\centering
\includegraphics[ bb=78 514 548 667, width=0.6\textwidth, clip]{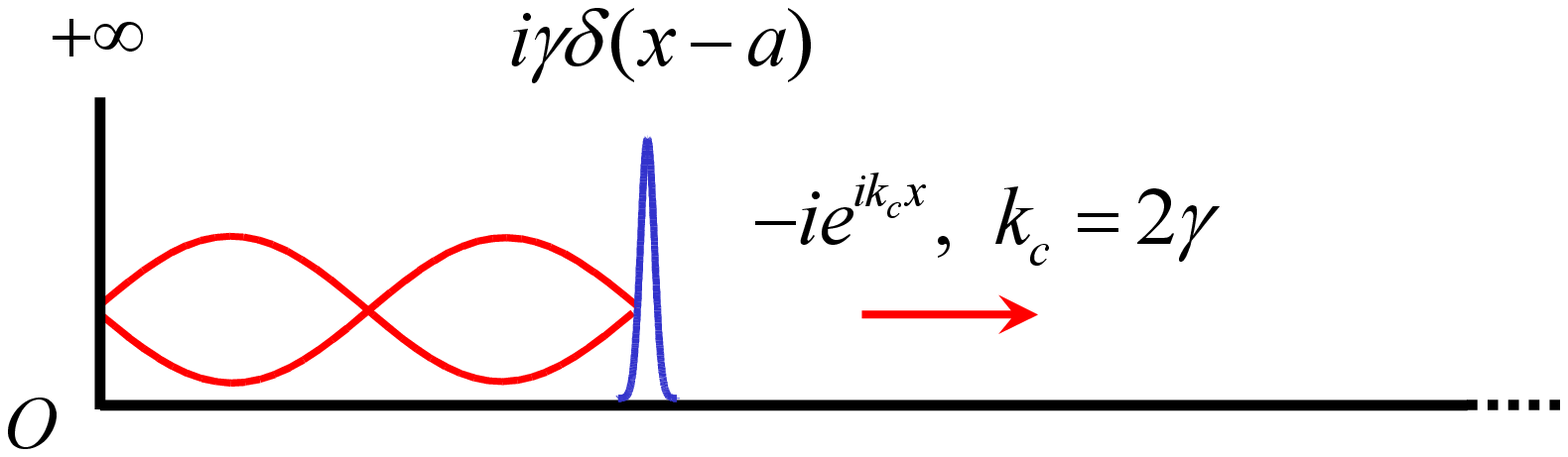}
\caption{(Color online) Schematic of the half-infinite non-Hermitian
Hamiltonian (\protect\ref{H_lasing}). The red line denotes the initial state
prepared in the cavity with wave vector $k_{c}=2\protect\gamma $ and $%
k_{c}a=\left( n+\frac{1}{2}\right) \protect\pi $, in which $n$ determines
the number of nodes of the initial standing wave.}
\label{fig4}
\end{figure}

\subsection{Solution}

SSs correspond to scattering states that behave like
zero-width resonances \cite{MostafazadehPRL2009}. They provide a
mathematical description of lasing or antilasing at the threshold \cite%
{MostafazadehPRA83,Longhiphysics,LonghiPRA1,LonghiPRA2}, which has stimulated a
series of theoretical and experimental studies concerning the laser
oscillators in optical systems \cite{Chong,Ge}. However, the previous studies
of these phenomena mainly focused on the systems possessing $\mathcal{PT%
}$ symmetry, which may restrict the structure of the optical medium, thereby
prohibiting the diverse experimental schemes. In this section, we
will propose a half-infinite non-Hermitian system without $\mathcal{PT}$
symmetry to simulate lasing-like behaviour. The Hamiltonian of the
composite system in question is given by
\begin{equation}
\mathcal{H}=-\frac{\partial ^{2}}{2\partial x^{2}}+V\left( x\right) ,
\label{H_lasing}
\end{equation}%
where%
\begin{equation}
V\left( x\right) =\left\{
\begin{array}{c}
i\gamma \delta \left( x-a\right) ,\text{ }\left( x>0\right) \\
\infty ,\text{ }\left( x\leqslant 0\right)%
\end{array}%
\right. ,
\end{equation}%
and $a$ is the length of the well. The composite scattering well consists
of an imaginary delta potential at $x=a$ and an infinite real
potential at $x=0$, which can be shown in Fig. 4. A particle in this
well is completely free, except at the two ends ($x=0$ and $x=a$), where an
infinite force prevents it from moving in the region of $x<0$. The infinite
potential at the left end can be deemed as the scattering center $B$
with the $r_{\mathrm{R}}^{B}=-1$ for right-incident wave with all
possible momentum $k$. The corresponding reflection amplitude of the
left-incident wave for the imaginary delta-potential can be given by
\begin{equation}
r_{\mathrm{L}}^{A}\left( k\right) =\frac{\gamma }{k-\gamma }.
\end{equation}%
Applying the condition (\ref{MC}) proposed in the Sec. 2.1 to this
composite continuous system, we can immediately obtain an SS without solving
Schr\"{o}dinger's equation as%
\begin{equation}
k_{c}=2\gamma ,\text{ }k_{c}a=\left( n+\frac{1}{2}\right) \pi ,
\label{condition}
\end{equation}%
which is determined by the strength of delta-potential $\gamma $ and the
length of the quantum well $a$. To obtain the corresponding eigenfunction,
we solve
\begin{equation}
\mathcal{H}\psi ^{k}\left( x\right) =\frac{k^{2}}{2}\psi ^{k}\left( x\right)
,
\end{equation}%
with boundary conditions
\begin{subequations}
\begin{eqnarray}
\psi ^{k}\left( 0\right)=0, \\
\psi ^{k}\left( a^{+}\right)=\psi ^{k}\left( a^{-}\right) , \\
\left[ \psi ^{k}\left( a^{+}\right) \right] ^{\prime }-\left[ \psi
^{k}\left( a^{-}\right) \right] ^{\prime } =2i\gamma \psi ^{k}\left(
a\right) ,
\end{eqnarray}%
where $\left[ ..\right] ^{\prime }$ denotes the derivative of the $\psi
^{k}\left( x\right) $ with respect to the coordinate $x$. This yields the
scattering solution
\end{subequations}
\begin{equation}
\psi ^{k}\left( x\right) =\left\{
\begin{array}{c}
\sin kx,\text{ }\left( 0<x<a\right) \\
\sin kx+\frac{2i\gamma }{k}\sin k\left( x-a\right) \sin ka,\text{ }\left(
x>a\right)%
\end{array}%
\right. .
\end{equation}%
Taking the Eq. (\ref{condition}) into $\psi ^{k}\left( x\right) $, we obtain
the following scattering solution at the SS%
\begin{equation}
\psi _{\text{SS}}\left( k_{c},x\right)\equiv\psi ^{k_{c}}\left( x\right)
=\left\{
\begin{array}{c}
\sin k_{c}x,\text{ }\left( 0<x<a\right) \\
-ie^{ik_{c}x},\text{ }\left( x>a\right)%
\end{array}%
\right. .
\end{equation}%
It shows that, outside the quantum well, the eigenfunction $\psi _{\text{SS}}\left( k_{c},x\right) $ represents a self-sustaining emission for $k_{c}>0$
(lasing) or a reflectionless absorption solution to $k_{c}<0$ (anti-lasing),
which is illustrated in Fig \ref{fig4}. In this sense, the quantum well can
be used to simulate a resonance chamber for for lasing where the SS of the system sets the lasing threshold. Based on these results, we  now characterize a lasing-like process through the time evolution of an initial state in the vicinity of an SS.
\begin{figure}[tbp]
\centering
\includegraphics[ bb=0 0 404 305, width=0.6\textwidth, clip]{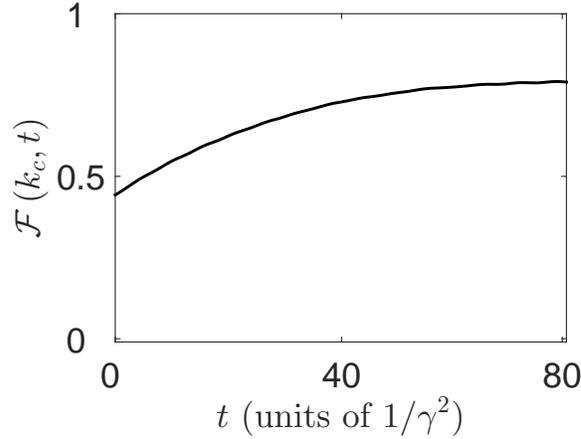}
\caption{(Color online) Plot of normalized fidelity $\mathcal{F}\left(
k_{c},t\right) $ as a function of time $t$. Parameter values are $\protect%
\gamma =1$, $k_{c}=2$, $a=10.25\protect\pi .$ It can be seen that $%
\mathcal{F}\left( k_{c},t\right) $ tends to be a steady value when $t$
increases beyond a given value, which is termed as relaxation-time $t_{f}$.}
\label{fig5}
\end{figure}

\subsection{Numerical simulation}

\begin{figure}[tbp]
\centering
\includegraphics[ bb=0 0 394 311, width=0.6\textwidth, clip]{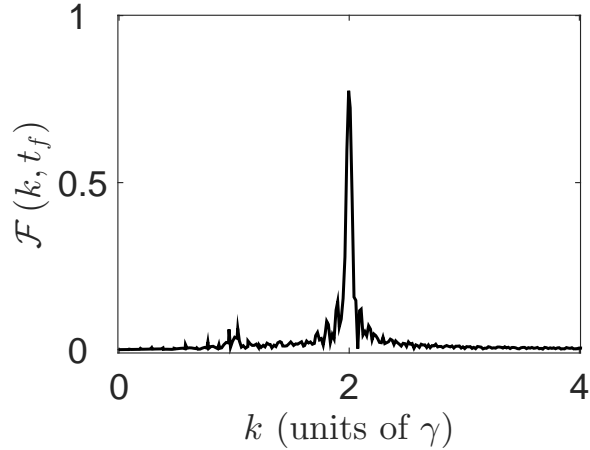}
\caption{(Color online) Probability distribution of $\Psi \left(
x,t_{f}\right) $ in the $k$ space. The system parameters are the same as
that of Fig. \ref{fig5}. One can see that the function of $\mathcal{F}\left(
k,t_{f}\right) $ are centered around $2$, which corresponds to an output
laser with narrow bandwidth.}
\label{fig6}
\end{figure}

We start with a sinusoidal profile as an initial state,%
\begin{equation}
\Psi \left( x,0\right) =\frac{1}{\sqrt{\Lambda }}\left\{
\begin{array}{c}
e^{ik_{c}x}-e^{-ik_{c}x},\text{ }\left( 0<x<a\right) \\
0,\text{ }\left( x>a\right)%
\end{array}%
\right. ,
\end{equation}%
where $\Lambda $ is the normalization factor. The distribution of the $\Psi
\left( x,0\right) $ in the cavity is the same as that of $\psi _{\text{%
SS}}\left( k_{c},x\right) $. As $a$ increases to infinity, the initial state
$\Psi \left( x,0\right) $ approaches the eigenstate $\psi _{\text{SS}%
}\left( k_{c},x\right) $ of the system. Therefore, the probability of the
evolved state $\Psi \left( x,t\right) $ driven by the Hamiltonian $\mathcal{H%
}$ in the cavity will not change and remains unity. For a finite-length cavity,
this result also indicates that the initial state $\Psi \left( x,0\right) $
mainly consists of $\psi _{\text{SS}}\left( k_{c},x\right) $ in the energy
space. Accordingly, one can assume that as time $t$ increases to infinity,
the evolved state $\Psi \left( x,t\right) $ will approach to the eigenstate $%
\psi _{\text{SS}}\left( k_{c},x\right) $, thereby mimicking the lasing
process. Towards this end, we first define the target state as $\psi _{\text{SS}%
}\left( k_{c},x\right) $, which constitutes ideal laser radiation emanating from the cavity. To measure the similarity between $\psi _{\text{SS}%
}\left( k_{c},x\right) $ and $\Psi \left( x,t\right) $, we introduce the
normalized fidelity as%
\begin{equation}
\mathcal{F}\left( k,t\right) \mathcal{=}\frac{\int \Psi ^{\ast }\left(
x,t\right) \psi ^{k}\left( x\right) \text{d}x}{\int \left\vert \Psi \left(
x,t\right) \right\vert ^{2}\text{d}x},
\end{equation}%
which also represents the purity of the laser output from the cavity. The
analytical result for this continuous half-infinite system is hardly
obtained, so we perform a numerical simulation to study the behaviour of
$\mathcal{F}\left( k_{c},t\right) $. The method used for the numerical simulation
is presented in Ref. \cite{Datta}. The key point in such a simulation is to discretise the continuous Hamiltonian into a lattice model. Fig. \ref{fig5} plots the normalized fidelity $\mathcal{F}\left( k_{c},t\right) $ as a function of time $t$. Note that, as $t$ increases, $\mathcal{F}\left( k_{c},t\right) $ tends to
be a steady value. For simplicity, we refer to the corresponding
time as the relaxation-time $t_{f}$. In addition, we focus on the
purity of the output wave in the $k$ space. When lasing, the output wave reflects the bandwidth of the laser, which is usually introduced to describe the monochromaticity of laser radiation. To quantitatively characterise the degree of monochromaticity, we plot the $\mathcal{F}\left( k,t_{f}\right) $ as a function of $k$ in Fig. \ref{fig6}, which shows that $\Psi \left( x,t_{f}\right) $ are centered mainly around $k_{c}=2\gamma $ representing a narrow-bandwidth laser. Owing to the complexity of
calculating the time evolution of the state $\Psi \left( x,0\right) $, we
cannot obtain the analytical solution for the bandwidth of the output wave.
However, according to the analysis above, when the initial state is prepared with $k_{c}=2\gamma ,$ $k_{c}a=\left( n+\frac{1}{2}\right) \pi $%
, one can envisage that a longer cavity will result in a narrower lasing bandwidth.

\section{Summary and discussion}

\label{sec_summary}

In summary, we show herein how to use the scattering solutions of two independent scattering centres to construct non-Hermitian composite systems with SSs. In essence, the left-incident solution for one scattering center and the right-incident solution for the other scattering center are matched through Eq. (\ref{MC}). The advantage of this method is that a composite system can be constructed with an SS at any desired energy by adjusting the parameters of the two scattering centres. Furthermore, we apply the proposed condition to a continuous and discrete non-Hermitian system. For such a system, we extend the condition to a system with multiple scattering centres, in which the wave vector associated with an SS equals the sum of the strengths of all the delta potentials. In addition, we construct a half-infinite non-Hermitian system to simulate lasing at the SS, which provides a clear physical application for SSs.

\section{Acknowledgements}

This work is supported by the National Natural Science Foundation of China
(Grants No. 11505126 and No. 11374163). X. Z. Zhang is also supported by the
Postdoctoral Science Foundation of China (Grant No. 2016M591055) and PhD
research start-up foundation of Tianjin Normal University under Grant No.
52XB1415.


\section*{References}

\begin{thebibliography}{99}
\bibitem{Moiseyev} Moiseyev N 1998 \emph{Phys. Rep.} \textbf{302} 212

\bibitem{Muga} Muga J G, Palao J P, Navarro B, and Egusquiza I L 2004 \emph{%
Phys. Rep.} \textbf{395} 357

\bibitem{Rotter} Rotter I 2009 \emph{J. Phys. A} \textbf{42}, 153001

\bibitem{Bender1998} Bender C M and Boettcher S 1998 \emph{Phys. Rev. Lett.}
\textbf{80} 5243

\bibitem{MostafazadehPRL2009} Mostafazadeh A 2009 \emph{Phys. Rev. Lett.}
\textbf{102} 220402

\bibitem{Andrianov} Andrianov A A, Cannata F, and Sokolov A V 2010 \emph{J. Math. Phys.} \textbf{51} 052104

\bibitem{MostafazadehJPA} Mostafazadeh A 2011 \emph{J. Phys. A} \textbf{44}
375302

\bibitem{MostafazadehPRA83} Mostafazadeh A 2011 \emph{Phys. Rev. A} \textbf{%
83} 045801

\bibitem{MostafazadehPRA2012} Mostafazadeh A and Rostamzadeh S 2012 \emph{%
Phys. Rev. A} \textbf{86} 022103

\bibitem{Kato} Kato T 1966 \textit{Perturbation Theory of Linear Operators}
(Berlin: Springer)

\bibitem{Berry} Berry M V 2004 \emph{Czech. J. Phys.} \textbf{54} 1039 (2004)

\bibitem{Kemp} Kemp R R D 1958 \emph{Can. J. Math.} \textbf{10} 447

\bibitem{Naimark} Naimark M A 1960 \emph{Am. Math. Soc. Transl.} \textbf{16}
103

\bibitem{Schwartz} Schwartz J 1960 \emph{Commun. Pure Appl. Math.} \textbf{13%
} 609

\bibitem{Ljance} Ljance V E 1967 \emph{Am. Math. Soc. Transl.} \textbf{60}
185

\bibitem{Samsonov} Samsonov B F 2005 \emph{J. Phys. A} \textbf{38} L397

\bibitem{Dehnavi} Mostafazadeh A and Mehri-Dehnavi H 2009 \emph{J. Phys. A}
\textbf{42} 125303

\bibitem{Heiss} Heiss W D 1994 \emph{Phys. Rep.} \textbf{242} 443

\bibitem{LonghiPRB} Longhi S 2009 \emph{Phys. Rev. B} \textbf{80} 165125

\bibitem{Longhiphysics} Longhi S 2010 \emph{Physics} \textbf{3} 61

\bibitem{LonghiPRA1} Longhi S 2010 \emph{Phys. Rev. A} \textbf{82} 031801(R)

\bibitem{LonghiPRA2} Longhi S 2011 \emph{Phys. Rev. A} \textbf{83} 055804

\bibitem{LonghiPRL} Longhi S 2011 \emph{Phys. Rev. Lett.} \textbf{107} 033901

\bibitem{Chong} Chong Y D, Ge L, Cao H, and Stone A D 2010 \emph{Phys. Rev.
Lett.} \textbf{105} 053901

\bibitem{Wan} Wan W, Chong Y, Ge L, Noh H, Stone A D, and Cao H 2011 \emph{%
Science} \textbf{331} 889

\bibitem{Ge} Chong Y D, Ge L, and Stone A D 2011 \emph{Phys. Rev. Lett.}
\textbf{106} 093902

\bibitem{Stone} Ge L, Chong Y D, Rotter S, T\"{u}reci H E, and Stone A D
2011 \emph{Phys. Rev. A} \textbf{84} 023820

\bibitem{Sarisaman} Mostafazadeh A, and Sarisaman 2013 M. \emph{Phys. Rev. A}
\textbf{87} 063834

\bibitem{Zhang} Zhang X Z, Jin L, and Song Z 2013 \emph{Phys. Rev. A}
\textbf{87} 042118

\bibitem{Li} Li G R, Zhang X Z, and Song Z 2014 \emph{Ann. Phys. (NY)}
\textbf{349} 288

\bibitem{ZhangG} Zhang G, Li X Q, Zhang X Z, and Song Z 2015 \emph{Phys.
Rev. A} \textbf{91} 012116

\bibitem{LiXQ} Li X Q, Zhang X Z, Zhang G, and Song Z 2015 \emph{Phys. Rev. A%
} \textbf{91} 032101

\bibitem{MusslimaniOL} El-Ganainy R, Makris K G, Christodoulides D N, and
Musslimani Z H 2007 \emph{Opt. Lett.} \textbf{32} 2632

\bibitem{Makris08} Makris K G, El-Ganainy R, Christodoulides D N, and
Musslimani Z H 2008 \emph{Phys. Rev. Lett.} \textbf{100} 103904

\bibitem{AGuo} Guo A, Salamo G J, Duchesne D, Morandotti R, Volatier-Ravat
M, Aimez V, Siviloglou G A and Christodoulides D N 2009 \emph{Phys. Rev.
Lett.} \textbf{103} 093902

\bibitem{CERuter} R\"{u}ter C E, Makris K G, El-Ganainy R, Christodoulides D
N, Segev M, and Kip D 2010 \emph{Nature Phys.} \textbf{6} 192

\bibitem{JPAAli} Mostafazadeh A 2006 \emph{J. Phys. A} \textbf{39} 13495

\bibitem{Datta} Datta S 1995 \emph{Electronic Transport in Mesoscopic Systems%
} (Cambridge: Cambridge Univ. Press, Cambridge)
\end{thebibliography}
\end{document}